\documentclass[aps,pre,twocolumn,groupedaddress]{revtex4-1}

\usepackage{amsmath}
\usepackage[dvipdfmx]{graphicx}% Include figure files
\usepackage[dvipdfmx]{color}% Include figure files
\usepackage[normalem]{ulem}

\usepackage{dcolumn}% Align table columns on decimal point
\usepackage{bm}% bold math

\begin{document}

\title{
Negative pressure in shear thickening bands of a dilatant fluid
}

%\author{Shin-ichiro Nagahiro$^1$, Hiizu Nakanishi$^2$ and Namiko Mitarai$^3$}
\author{Shin-ichiro Nagahiro$^1$ and Hiizu Nakanishi$^2$}
\affiliation{
$^1$Department of Mechanical Engineering, 
Sendai National College of Technology,
Miyagi 981-1239, Japan
}

\affiliation{
$^2$Department of Physics, Kyushu University 33, Fukuoka 
{819-0395}, Japan}

\date{\today}

\begin{abstract}
We perform experiments and numerical simulations to investigate spatial
distribution of pressure in a sheared dilatant fluid of the
Taylor-Couette flow under a constant external shear stress.  In a
certain range of shear stress, the flow undergoes the shear thickening
oscillation around 20 Hz. We find that, during the oscillation, a
localized thickened band rotates around the axis with the flow.
{Based upon experiments and numerical simulations, we show that} a
major part of the thickened band is under negative pressure {even
in the case of discontinuous shear thickening}, which indicates that the
thickening is caused by Reynolds dilatancy; the dilatancy causes the
negative pressure in interstitial fluid, which generates contact
structure in the granular medium.{, then frictional resistance
hinders rearrangement of the structure and solidifies the medium.}
\end{abstract}


%Use showkeys class option if keyword display desired
%\keywords{}

\maketitle
%----------------------------------------------------------------------

\section{Introduction}
A fluid with suspended particles have an apparent viscosity %which is 
different from that of the medium fluid.  If the total volume of
particles is much smaller than that of fluid, the suspension behaves
as a Newtonian fluid {with the viscosity given by} well known Einstein's
viscosity formula \cite{landau}.  In the case of high volume fraction,
the fluid exhibits non-Newtonian behaviors.  The viscosity decreases
with increasing shear rate (shear thinning)\cite{Rutgers} in many
materials such as mud, paint, and ketchup.  Shear thickening
behaviors, i.e., increasing viscosity with shear rate, are also observed
in some other suspensions.  In the extreme case such as the dense suspension
of starch particles in water, the fluid almost solidifies under shear
stress; the viscosity increases discontinuously by orders of magnitude
\cite{wagner,fall,jaeger}. They are often called ``dilatant fluid'' due
to apparent analogy to Reynolds dilatancy of granular media \cite{freu},
and their fascinating and unintuitive behaviors are popular subjects for
science demonstrations \cite{merkt,ebata,stefan2,isa,naga}.

The mechanism of discontinuous shear thickening (DST) is still under
debate. A promising explanation is related to the dilatancy and
jamming \cite{cate}. As stated in the Reynolds principle of dilatancy, 
dense granular media must dilate when they deform. If a suspension is
confined, the dilation leads to jamming and then shear stress abruptly
increases.
It is numerically \cite{otsuki} and experimentally \cite{dapeng} 
demonstrated that contact friction between particles is important for DST 
in a shear flow of dry granular systems.  
The frictional contacts is  found to be essential for DST 
also for the hard-sphere suspensions \cite{seto_lett,seto_full, nicolas}. 
Lin {\it et al.} recently found that even in continuous shear thickening
the contact forces between particles dominate hydrodynamic
interactions \cite{neil}.
With these results, it is argued that DST is a consequence of jamming
caused by dilatancy in a medium of frictional particles.
%

%---------- Shear Thickening Oscillation ---

{
It has been known that the uniform steady shear flow is unstable for
shear thickening media and noisy fluctuation or periodic oscillation
under a constant external shear stress have been reported
\cite{laun,lootens1,naga,larsen-2014}.  Such oscillation is a general
feature of a shear thickening fluid with the S-shaped flow curve
\cite{bashkirtseva-2010}, and may be interpreted as {\it shear
thickening oscillation}, i.e., a periodic alternation between the
thickened and thinned state of the media under a constant external
stress that is in the unstable branch of the flow curve.  This
oscillation has been predicted by a fluid dynamics model for shear
thickening media \cite{nakanishi1,nakanishi2}, and observed
experimentally in shear flows of macroscopic width of several
centimeters \cite{naga}.  The frequency of the oscillation is expected
to depend weakly on the width of the flow if one assume spatially
uniform thickening, but spatial inhomogeneity develops quickly and
usually only one thickened band remains after initial transient.  Even
with such spatial inhomogeneity, a clear
oscillation around 20 Hz appears in the experiment with a macroscopic
flow width around 5 cm, and its frequency hardly depends on either the
flow width or the external stress \cite{naga}.  There are also
experiments that report similar oscillations in the shear flows with
narrower width, typically 1 mm or less than a hundred of
particle diameters\cite{laun,lootens1,larsen-2014}, in which case the
particle discreteness may play some role.
}

%---------------------------------------------

In this {work}, we investigate the Taylor-Couette flow of a dilatant
fluid by experiments and numerical simulations.  We measure the
off-center force on the axis and the local pressure at the wall of the
outer cylinder. We also perform numerical simulations for a three
dimensional system using a fluid dynamics model developed by the authors
\cite{nakanishi1, nakanishi2}.  The comparison of the simulation results
with the experimental data reveal the spatial distribution of pressure
and viscosity in the flow. {We find that the thickening region 
strongly localizes, and forms two types of thickening bands, which 
distinctly {have} positive and negative pressure. It is remarkable that, 
even in the DST regime, }the dominant thickening bands extend 
in the stretching direction under 
negative pressure, namely, the system jams due to tensile stress. Our 
local pressure measurement uncovers that the negative pressure is 
limited by the Laplace pressure, suggesting that the jamming under 
the tensile stress is sustained by the interstitial fluid. 

%In experiments in the parallel- or in the cone-plate
%geometry on a stress-controlled rheometer, suspensions dilates at the
%onset of shear thickening, exerting a positive normal force
%on the plates \cite{lootens1,fall,ericjaegar}.  
%The contact force network in the granular media is supposed to
%span in the compressing direction.

%--------------------------------------------------------------
\begin{figure}[b]
\begin{center}
\includegraphics[width=7cm, keepaspectratio]{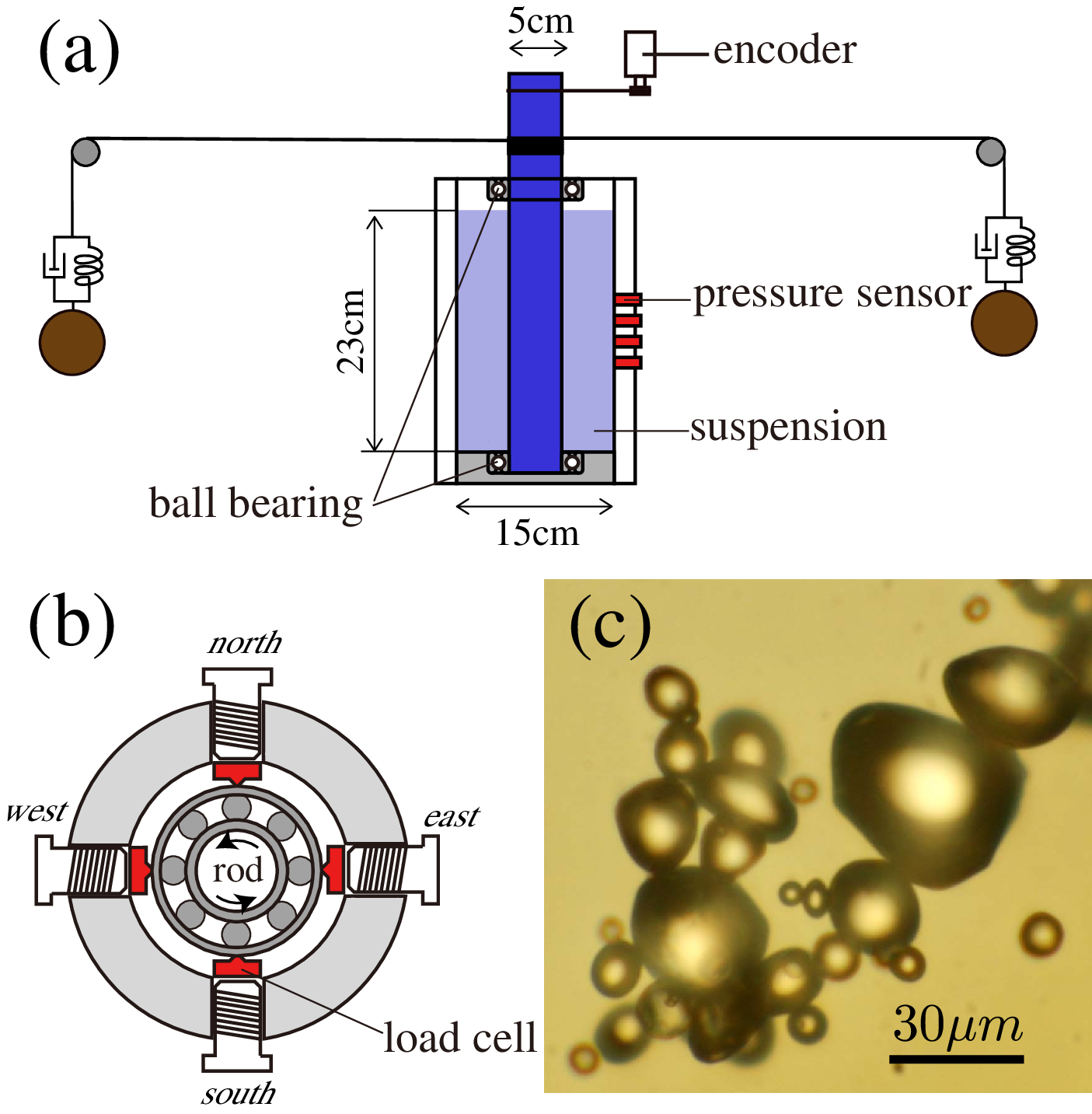}
\caption{(a) Schematic illustration of the experimental setup.  A
cylindrical container with a rotatable center rod is filled with
starch-water mixture.  A pair of weights are hung through steel
wires wound around the center rod to give constant torque to the
rod in CCW direction.  The flow width
$h$, i.e.  the gap between the cylinder and the rod, is $5$ cm.  (b) Top
view of the upper fixed end of the center rod.  The ball bearing
is supported at four points by load cells to measure
 the force acting on the rod.  (c)
Micrograph of the potato-starch particles.}  \label{fig:expset}
\end{center}
\end{figure}
%-------------------------------
\begin{figure}
%\centerline{\includegraphics[width=8cm]{./Flow-Curve.eps}} 
\centerline{\includegraphics[width=8cm]{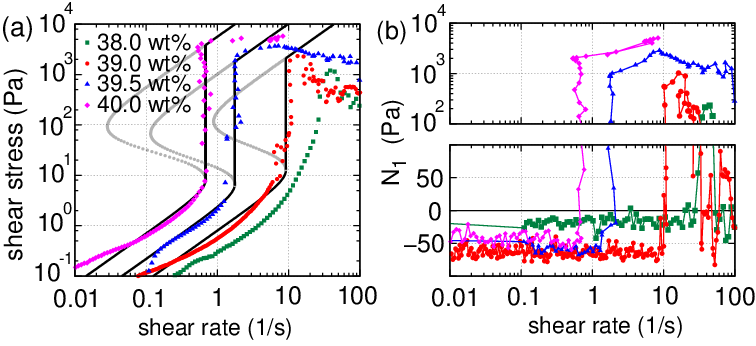}} 
\caption{{Flow curves (a) and the first normal stress difference
 $N_1$ (b) of the potatostarch suspension density-matched by CeCl.  The data
 are obtained by a cone-plate rheometer (Haake Mars) with the cone angle
 82$^\circ$, the gap size 140 $\mu$m, and the cone diameter 35 mm.  The
 data beyond DST are not meaningful because samples are fractured upon
 solidification.  The solid curves are those for the model used in the
 simulations {(Sec. \ref{sec:model})}. } } \label{Flow-Curve}
\end{figure}
%--------------------------------------------------------------
\begin{figure}[b]
\begin{center}
\includegraphics[width=8cm, keepaspectratio]{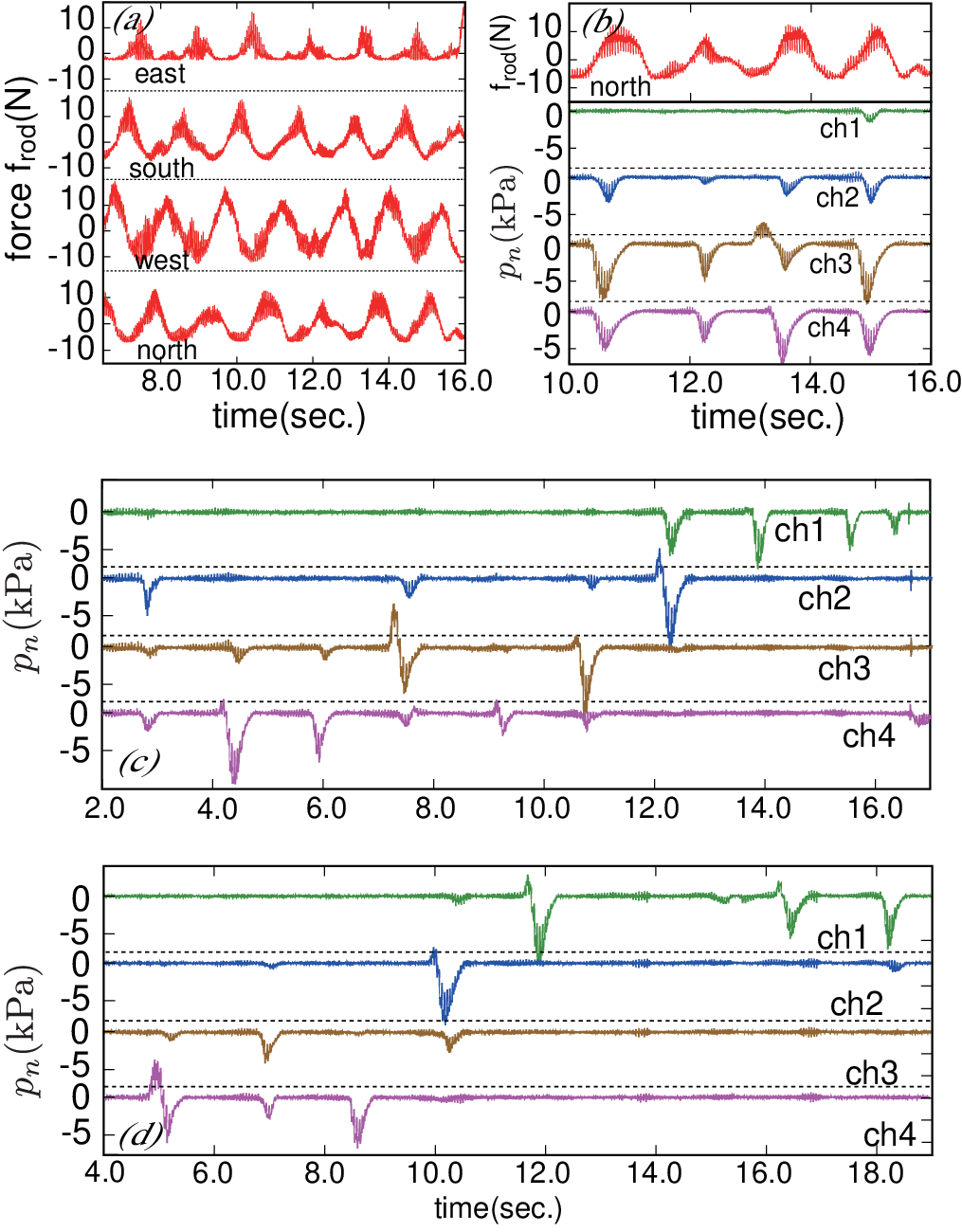}
\caption{Time evolutions of the off-center force on the center
    rod and the normal {pressure}  at the wall of the outer cylinder.  The
    hydrostatic pressure is subtracted from the normal {pressures}.  (a)
  Off-center forces for $S_e=1.46$ kPa. (b) The off-center force on the
  north load cell and the normal {pressures} for $S_e=1.46$ kPa plotted with the
  common time axis.  (c,d) Normal {pressures} for $S_e=0.84$ kPa and for
  $S_e=1.26$ kPa.  The suspension is a 39 wt$\%$ mixture of potato
  starch and an aqueous solution of CsCl with a density of {1.75} ${\rm
    g/cm^3}$.}
\label{fig:press_offf}
\end{center}
\end{figure}
%--------------------------------------------------------------

\section{Experiment}

%-----------------------------------------------------------------
\subsection{Systems}
  
Our experimental setup is shown in Fig.\ref{fig:expset}(a).  The
container consists of an acrylic outer cylinder (15 cm inner
diameter) and an acrylic center rod (5 cm diameter) with an
aluminum base plate and a lid; 
the gap $h$ between the outer cylinder and the central rod is
$h=5$ cm.
{The fluid fills the container up to $\ell=23\,{\rm cm}$ from the
bottom with its surface open to the air.}
The center rod is supported with a couple of ball bearings: one
embedded in the base plate and the other in the lid of the container.
Constant torque in the counterclockwise(CCW) direction is applied
on the center rod by a pair of weights on the opposite sides
through steel wires wound around the rod.  The weights are
hung through  spring-dumper systems to reduce the
vibration transmitted  from the rod.
The both weights are the same so that no net force should be
applied on the rod.
We use the weights in the range of $0.20\sim 4.0{0}\; {\rm kg}$,
which corresponds to the external shear stress $S_e=0.08\sim1.68\;
{\rm kPa}$ at the rod surface.  In order to enforce the no-slip
condition, the surface is lined with water-proof sand paper.

The off-center force on the center rod from the suspension is 
measured by four load cells (Kyowa LMB-A-100N), which support
the upper ball bearing at four points as shown in Fig \ref{fig:expset}(b).
To measure both negative and positive force, the load cells are 
pressed by screws
and the zero-point of the load cells is set
when the rod is stationary before each experiment.
We label them as ``north", ``south", ``east" and ``west" by their directions.
Note that precise calibration in the off-center force measurements is
difficult due to the friction at the contacts between a load cell and
the ball bearing.

We also measure the normal pressure $p_n$ at the surface of
the outer cylinder by four pressure sensors (Kyowa PGM-G-02KG).  They
are located to the ``north" of the center and aligned along the axial
direction at intervals of 2 cm with the shallowest one located at 7.5 cm
below the fluid surface; they are labelled as ``ch1'', ``ch2'', ``ch3''
and ``ch4'' from the top to the bottom.   The normal pressure
obtained by these sensors is the sum of the pressures by both the
interstitial liquid and the particles%\footnote{{The positive (negative)
%pressure {means} compressive (tensile) stress in the medium between the 
%center rod and outer cylinder.}}
.

We use the potatostarch particles (Hokuren) of irregular shape with
their sizes distributed over the range of $5\sim 30\,{\rm \mu m}$
[Fig.\ref{fig:expset}(c)].
The powder is dried for 24 hours at $60^\circ \rm C$ and 35$\%$
humidity, and then the water-starch mixture is prepared with the density
matched aqueous solution ({$\rho=1.75\,{\rm g/cm^3}$}) of cesium chloride
(CsCl).

%----------------------------------
\subsection{Results}

{
\paragraph*{Rheology of the media:}
First of all, the flow curves and the normal stress difference $N_1$
measured by a cone-plate rheometer for the water-starch mixtures are
shown in Fig.\ref{Flow-Curve} to present rheological properties of the
media that we are going to study.
DST is clearly observed except for the 38 wt\% mixture.  The data
presented in the rest of this report are for the 39 wt\% mixture.  In
our experimental setup of Fig.\ref{fig:expset}(a), the shear thickening
oscillation of 20 Hz is observed for $S_e\gtrsim 0.1$ kPa, as will be
presented below. This is in the unstable branch of the flow curve in
Fig.\ref{Flow-Curve}(a), thus is consistent with our interpretation that
the 20 Hz oscillation is the shear thickening oscillation.
The solid curves and gray plots in {Fig.\ref{Flow-Curve}(a)} are the flow
curves upon increasing shear rate and the unstable branch, respectively,
by the model used in the simulations with the parameters listed in Table
\ref{parameters}.  One can see that the rheology of the media is well
reproduced by the model. Detailed description of the model is given in
Sec.IIIA.
The normal stress difference $N_1$ shown in {Fig.\ref{Flow-Curve}(b)} are
negative for small shear rate, but upon DST they jump to large positive
values in agreement with previous results\cite{royer-2016}.
}

%-----------------------------
\paragraph*{Measurement by Taylor-Couette cell:}
Figure \ref{fig:press_offf} shows the typical time developments of the
off-center force $f_{\rm rod}$ on the center rod and the normal {pressure}
{$p_n$} at the outer cylinder during the shear thickening
oscillation.  In Fig.\ref{fig:press_offf}(a), the results from the four
load cells at east, south, west, and north are shown for $S_e=1.46$
kPa\footnote{
The time averages of the forces are subtracted from the measured
  data because the zero of $f_{\rm rod}$ shifts unpredictably during
  the experiments due to friction force at the contact of the load
  cells. However, the subtracted value is smaller than the amplitude
  of $f_{\rm rod}$ ($\pm 4$N at the largest), thus we can regard the
  positive (negative) $f_{\rm rod}$ corresponds to compressive
  (pulling) force on a load cell. }.
The curves are roughly sinusoidal shape with the period $\tau_b\simeq 1.4$ s 
overlaid by the characteristic oscillation of shear thickening
with the period $\tau_{\rm sto}\simeq 0.05$ s.
The time shifts between the plots for the neighboring load
cells are $1/4$ of their periods, indicating that the direction of the
off-center force rotates with the flow to the CCW direction. 

Figure \ref{fig:press_offf}(b) shows the temporal variations of the
off-center force $f_{\rm rod}$ on the north load cell along with the
normal {pressures} {$p_n$} measured by the four pressure sensors
on the outer cylinder wall in the north. Note that the hydrostatic
pressure is subtracted from the normal {pressures}.
We observe periodic negative  pulses in {$p_n$} with the  width
$\tau_p\simeq 0.6$ s. They are  slightly ahead of
the peaks in the ``north'' component of $f_{\rm rod}$.
One may notice that the negative pulses are occasionally preceded by
relatively weak positive pulses.  It is also notable that many of the
pulses are detected only by a couple of sensors, which reveals that the
pressure fluctuation is localized in a region of a few centimeters
thickness in the axial direction.
These features are also seen in the {$p_n$} data for $S_e=$1.05 and
0.84 kPa in Fig.\ref{fig:press_offf}(c,d).
%

%--------------------------------------------------------------
\begin{figure}%[t]
\begin{center}
\includegraphics[width=8.5cm]{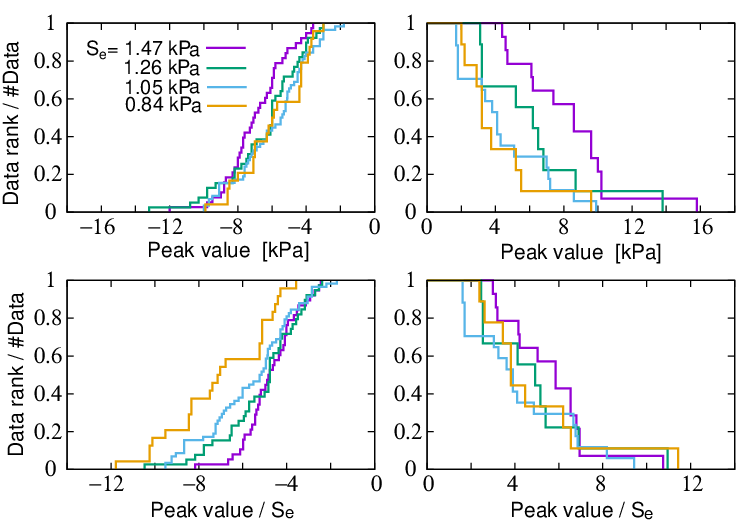}
\end{center}
\caption{ Normalized rank plots, i.e. the integrated distributions, of
the peak values of the normal {pressure $p_n$} for the negative
pulses (left panels) and the positive pulses (right panels) for various
external shear stresses $S_e$.
The data ranks are plotted against the bare peak values in the upper plots
while the peak values are scaled by the external stress $S_e$ in the
lower plots.
The plots for the various $S_e$ overlap with each other 
better in the bare plots for the negative pulse while
data scatter is smaller in the scaled plots for the positive pulse.
} \label{fig:peakp}
\end{figure}
%--------------------------------------------------------------

\paragraph*{Positive and negative pressure pulses: }
Figure \ref{fig:peakp} shows the integrated distributions of the peak
values of {$p_n$} for the positive and the negative
pressure pulses for various external shear stresses.  The integrated
distributions are plotted against the {peak values} in the
upper panels, while the horizontal axes are scaled with the external
shear stresses $S_e$ in the lower panels.  For the negative pulse, the
plots in the upper panel collapse along a single common curve better
than those in the lower panel.  As for the positive pulse, better
collapse is obtained by the scaled plots in the lower panel.

The better collapse for the scaled data of the positive pulse means that
the pressure in the positive pulse is proportional to the externally
applied shear stress.  This result may corresponds to the linear
correlation between the first normal stress difference $N_1$ and
external shear stress reported by Lootens et. al \cite{lootens1}, and in
accords with the picture of the jamming mechanism.

In contrast to the positive pulse, the peak pressure distribution of the
negative pulse is almost independent of $S_e$, and the maximum {negative} pressure
in the distributions are {$-14\sim -10$} kPa.  These values are close to
the Laplace pressure $-2\gamma/R$, which gives $-10\,{\rm kPa}$ for the
average particle size of our potatostarch particles ($15\,{\rm \mu m}$)
and the surface tension of water ($73$ mN/m).

%----------------------------------------------------------------
\begin{figure*}%[b]
\begin{center}
\includegraphics[width=14cm,keepaspectratio]{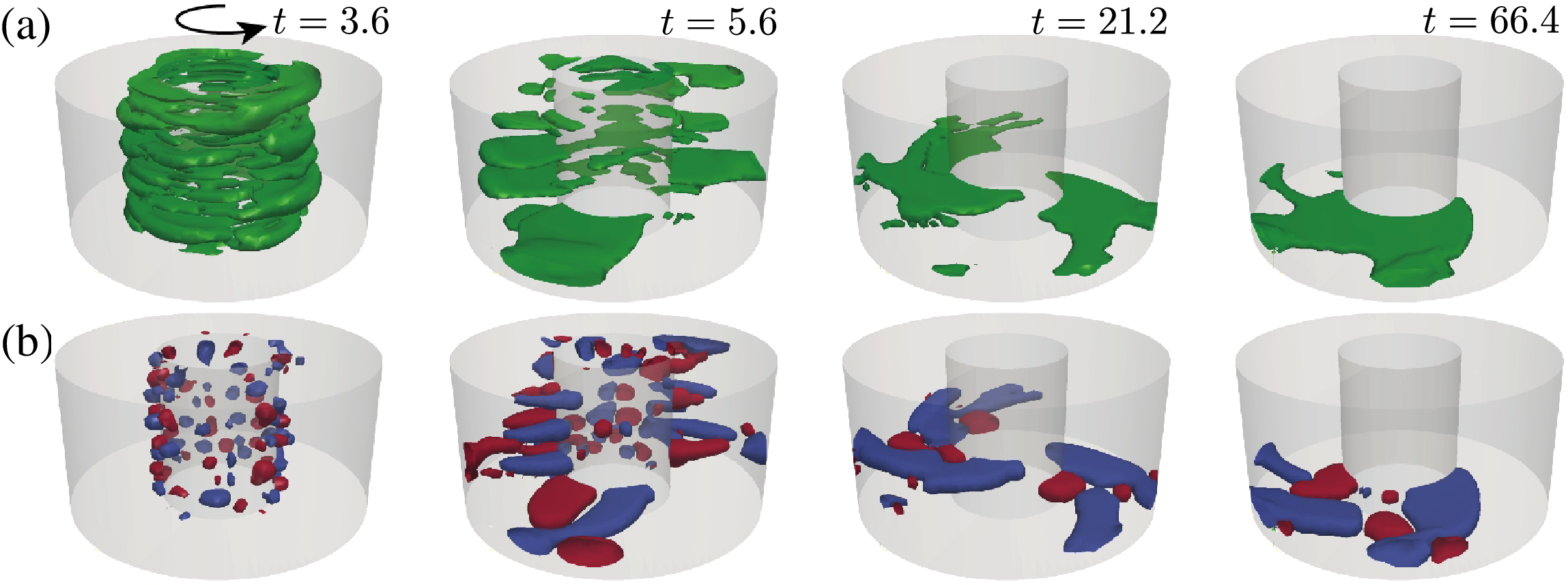}
\caption{Snapshots of the numerical simulation of Taylor-Couette flow
{from  the uniformly relaxed initial state
by} a phenomenological fluid dynamics model of a dilatant fluid
\cite{nakanishi1, nakanishi2}.  The depth of the flow is $\ell
=2.6$, the gap between the two cylinders is $h=1.5$, and
the external stress on the surface of center rod is $S_e=1.5$. The
arrow indicates the direction of rotation.  (a) Isosurface of viscosity
for $\eta=2$. (b) Isosurface of isotropic pressure for $p=1$ (red) and $p=-1$
(blue). The unit system is defined {in the text}.}
\label{fig:numerics}
\end{center}
\end{figure*}
%--------------------------------------------------------------

\section{Numerical Simulation}

\subsection{Model}\label{sec:model}
We performed three-dimensional (3-$d$) simulations for the
Taylor-Couette flow using a fluid dynamic model of a dilatant fluid
\cite{nakanishi1,nakanishi2}.  The model is based on the incompressible
Navier-Stokes equation
\begin{align}
 \rho{Dv_i\over Dt} 
& = {\partial\over\partial x_j}\big(-p\delta_{i,j}+\sigma_{i,j}\big)
\end{align}
with the stress and the strain rate tensors
\begin{align}
 \sigma_{i,j} &=\eta(\phi)\dot{\gamma}_{i,j},
\\
\dot{\gamma}_{i,j} 
&\equiv{\partial v_i\over\partial x_j}+{\partial v_j\over\partial x_i},
\end{align}
where Lagrange derivative is denoted by
\[
 {D\;\over Dt} \equiv {\partial\;\over\partial t} +
 v_i{\partial\;\over\partial x_i},
\]
and Einstein rule for the summation is employed.
The scalar field $\phi$ is
a phenomenological parameter
supposed to represent the internal structure of the
medium, such as the contact number of grains, but
 {\em not}  a conserved quantity like the volume fraction.
The pressure $p$ is determined by the incompressible condition
\begin{equation}
 {\partial v_i\over\partial x_i} = 0.
\end{equation}

The viscosity of the medium $\eta$ is a function of $\phi$, and local
value of $\phi(\bm r)$ is driven by the shear deformation to the value
$\phi_*$ determined by the local shear stress $S$ of the medium,
\begin{align}
 r{D\phi(\bm r)\over Dt} & = -\dot\gamma\big(\phi(\bm r)-\phi_*(S)\big),
\label{eq-phi}
\end{align}
where $r$ is a dimensionless parameter and
\begin{align}
\dot\gamma^2  &\equiv {1\over 2} \dot\gamma_{i,j}\dot\gamma_{j,i},
\\
S^2 & \equiv {1\over 2} \sigma_{i,j}\sigma_{j,i}.
\end{align}
Note that in Eq.(\ref{eq-phi}) the time derivative is assumed to be
proportional to the strain rate, which means that the change of $\phi$
is driven by the strain and the dimensionless parameter $r$ represents
the strain scale that drives $\phi$.
We also like to remark that our model is not for the shear rate
thickening, but for the shear stress thickening because the viscosity is
determined by the shear stress $S$ through the function $\phi_*(S)$.

%-------------------
\begin{table*}
\centerline{\tabcolsep=0.4cm
\begin{tabular}{c|cccc|cc}
\hline
concentration [wt\%]    &  $\phi_M$   &  $S_0$ [Pa] 
  &  $\eta_0$ [Pa$\cdot$s]   &  $A$ 
& $\tau_0$ [s]  &  $\ell_0$ [cm]  \\\hline\hline
 40.0   &  0.87 & 10.0 & 7.0 & 1 & 0.7 & 5.3 \\
 39.5 &  0.86 & 8.0  & 2.2 & 1 & 0.28 & 1.9 \\
 39.0 &  0.85 & 15.0 & 0.8 & 1 & 0.53& 0.49\\
\hline
\end{tabular}}
\caption{Parameters used for the plots (solid and dotted lines) by the model in
Fig.\ref{Flow-Curve}.
The values for $\tau_0$ and $\ell_0$ are estimated for
{$\rho = 1.75\times 10^3 {\rm kg/m^3}$}.}
  \label{parameters}
\end{table*}
%-------------------

The functional forms of $\phi_*(S)$ and $\eta(\phi)$ are chosen so
that the model can simulate behaviors of the dilatant fluid.
Employing simple functional forms
\begin{align}
 \phi_*(S) & \equiv \phi_M {(S/S_0)^2\over 1+(S/S_0)^2},
 \\
 \eta(\phi) & \equiv \eta_0 \exp\left[A{\phi\over 1-\phi}\right],
\end{align}
we have demonstrated that the model reproduces most of the
characteristic behaviors of the dilatant
fluids\cite{nakanishi1,nakanishi2,naga}.

Assuming the steady uniform shear flow under an external shear stress
$S$, the shear rate $\dot\gamma$ is given by
\begin{equation}
\dot\gamma = {S\over\eta\left[\phi_*(S)\right]}.
\end{equation}
This gives the S-shaped flow curve, which have the unstable branch
between the low and the high stress stable branches.  In Fig.\ref{Flow-Curve},
this relation is compared with the experimental flow curves
obtained by increasing stress.  The solid curves for the model are drawn
by assuming that the stress jump from the low stress branch to the high
stress branch at the end of the low stress branch.
The model parameters used in Fig.\ref{Flow-Curve} are listed in Table 1.

In this model, the system is characterized by the dimensional material
parameters, $\rho$, $S_0$, $\eta_0$, and the dimensionless model
parameters, $\phi_M$, $r$, $A$.  As for the dimensional parameters, we
can define the unit system where $\eta_0 = S_0 = \rho = 1$, then, the
time, length, and mass are measured by the units
\begin{equation}
 \tau_0 \equiv {\eta_0\over S_0}, \quad
  \ell_0 \equiv \sqrt{{\eta_0\over\rho}\tau_0}, \quad
  m_0 \equiv \rho \ell_0^3,
\label{unit}
\end{equation}
respectively.
As for the  dimensionless model parameters, we took
\[
 \phi_M=0.85, \quad r=0.1,\quad A=1
\]
in the 3-$d$ simulations.
The time and the length units given by Eq.(\ref{unit}) are listed also
in Table \ref{parameters} for the present system with
{$\rho = 1.75\times 10^3 {\rm kg/m^3}$}.

We have already demonstrated that the model reproduces basic
properties of a dilatant fluids such as DST, the hysteresis upon
changing shear rate, or instantaneous solidification by an external
impact\cite{nakanishi1}.  The shear thickening oscillation had been
predicted by this model and was experimentally observed as
had been predicted\cite{naga}.
%--------------------------------

We employ the Highly Simplified Marker and Cell (HSMAC)
algorithm\cite{hirt} for numerical simulations.  As for the boundary
conditions at the cylinder and the rod surfaces, we emulate the ones
in our experiment, i.e. the no-slip fixed boundary at the outer
cylinder and the no-slip boundary with the rotating center rod,
which rotates so as to give an average shear stress on the
surface equal to the applied shear stress $S_e$; we ignore the mass
of the rod.
As for the boundaries in the rotating axis direction, however, we
employ the periodic boundary condition for simplicity.  
We set diameter of the center rod
$d=2$, the flow width $h=1.5$ and the fluid depth $\ell=2.6$ {by the
units (\ref{unit})}.  With the parameters
for the present medium, they are comparable with the ones for our
experimental set up.

%---------------------------------
\subsection{Results}

A uniform steady flow is unstable for the external stress $S_e$ beyond a
certain value.  
{ Figure \ref{fig:numerics} show the system evolution during the
  first 2.5 rotations of the central rod under the external stress
  $S_e=1.5$, which is in the unstable regime.  The system is initially
  in the uniform relaxed state.  The upper panels show the
  isosurfaces at $\eta=2$ (green) and the lower panels show those for
  the isotropic pressure $p=+1$ (red) and for $p=-1$ (blue).  } The
results show that the viscosity and pressure distribution are not
cylindrically symmetric.  High viscosity regions {initially appear
  near the center rod ($t\lesssim 3.6$), then gradually merge into a few
  flat regions ($t\lesssim 21.2$)}, and eventually form a single
fan-shape thickening band, in which positive and negative pressure
segments extend in different directions {($t=66.4$)}, 
i.e. the compressing and the stretching
directions in the shear flow caused by the rotating center rod. {The
  thickening band with the positive and negative segments rotates
  slowly with the center rod.}  As a result, the positive pressure
segments always go ahead of the negative pressure segments at the
surface of the outer cylinder.  This is in agreement with the results
of our normal pressure measurement, where the positive pulses precede
the negative pulses.

{In the simulations, the localized thickening band shows the shear
thickening oscillation whose period is much faster than that of the
rotation.  This oscillation in the simulation should correspond with the
20 Hz oscillation in our observation.
The basic mechanism of the oscillation is the
same with that in the 2-$d$ simple shear flow\cite{nakanishi2};}
in the S-shaped flow curve, there exists a range of
the shear stress where the steady flow is unstable; under an external
shear stress in the unstable range, no steady shear flow is possible
and the system oscillates between the thickened and the relaxed
states.  In the shear stress thickening, the spatially uniform flow in
the directions perpendicular to the shearing is intrinsically unstable
because localized thickened bands can take most of the external
stress, leaving the rest of the medium in the unthickened state under
low stress.

By the simulations of the Taylor-Couette flow, we can observe the
dynamics of the thickened band; starting from the uniform relaxed state,
the thickening regions first appear {in the initial transient time} 
near the inner rod, where the shear rate is large.  Then, some of the
regions extend outwards, but eventually only one of them remains and
reaches the outer cylinder.  When the band reaches the outer cylinder,
the flow decelerates.  Then, it starts flowing again because the
external stress is not large enough to keep the whole band in the
thickened state.  As the system starts flowing, the thickened band
breaks in the outer regions, then the flow accelerates further until the
shear stress causes thickening in the broken part of the band.
{During the oscillation cycle, only a part of thickening band
disappears, but the rest remains thickened.}

%------------------------------------------------------------

\section{Discussions}
The results of our numerical simulations and experiments are consistent
and show that the thickening band distinctly has two types of
segments: the positive pressure segment extending along the compressive
direction of the shear flow and the negative pressure segment along the
tensile direction. 
{The remarkable finding is that the negative pressure
segment of the shear thickening domain is dominant in volume
when the medium exhibits the shear thickening oscillation 
{in the Taylor-Couette flow.}}

%The remarkable finding is that the negative pressure
%segment is dominant in the shear thickening domain of 
%the medium in the Taylor-Couette flow.

%---------------------------
{The pressure observed by the sensors is the total pressure from
both the fluid and the particles, but it should be noted that the
negative contribution to the pressure only comes from the fluid because
the particles cannot exert tensile force on the sensors.  This may
entail to the observed differences} in the external shear stress
dependence of the pressure in the regions of positive and negative
pressure as shown in Fig. \ref{fig:peakp}; the pressure in the positive
domains increases linearly with the external shear stress $S_e$, while
the negative pressure does not seem to depend on $S_e$, {but remains} below
the Laplace pressure.  Such linear dependence of the positive pressure
has been reported also in Ref.\cite{lootens1}, and {suggests that
the positive pressure is dominated by the particle pressure, that
propagates along force chains in the jammed granular medium.}
As for the negative pressure, it should be caused by the fluid in the
interstitial space that {tends} to expand upon deformation of the medium
due to Reynolds dilatancy. In the experiment, thickening band does not
reach to the fluid-air interface at the top. Therefore, 
the fact that the negative pressure is limited by the Laplace pressure 
indicates that there exist the fluid-air interfaces possibly at the 
surface of micro-bubbles in the medium.

The shear thickening due to jamming in the negative pressure region has
not been discussed in the literature, but it is natural for the granular
medium with friction because the negative pressure in the interstitial
fluid should increase contacts among the granular particles and the
friction resists against rearrangement of the contact structure in the
medium.  It should be also noted that the shear thickening in the
tensile deformation is easily observed in a simple demonstration by just
pouring the starch-water mixture out of a cup, and the effect of shear
thickening on drop formation in a granular suspension has been
studied\cite{pan}.

%----------------------------

Although our experiments clearly show the existence of the negative
pressure regions, it has not been reported in the literature
\footnote{The weak negative pressure has been reported for a low
concentrated medium in the continuum shear thickening regime in
ref.\cite{lootens1}}; some experiments report only positive pressure
when DST occurs\cite{fall,lootens1,ericjaegar}.  These experiments,
however, do not observe the spatial variation of the pressure, but
measure only total force on the upper plate of a cone-plate or
plate-plate type rheometer, using small samples.  In such measurement,
the effect of the negative pressure may be hidden by the large positive
pressure under strong external shear stress in the case where the value
of the negative pressure is limited, even though the size of the
negative pressure region is not small in comparison with that of the
positive pressure.

%-------------------------

In conclusion, our experiments and numerical simulations show that, 
the negative pressure segment along the tensile direction is dominant 
in the shear thickening band of a dilatant fluid.  
%1
The negative pressure in the thickening bands
indicates that the thickening is caused by Reynolds dilatancy; the
negative pressure caused by the dilatancy generates contact structure in
the granular medium, {and the solidification of the medium is due to the
frictional resistance against the rearrangement of the structure.}  

%-------------------------
\begin{acknowledgements}
We thank Hisao Hayakawa and {Masahiko Okumura} for discussions, 
Madoka Nakayama for preparing
the micrograph of the particles, Takenobu Kato and Keiichi Sato for
technical assistance. This work is supported by MEXT KAKENHI grant
number 15K05223.
\end{acknowledgements}

\bibliography{citations}

\end{document}